\documentclass[galaxies,article,accept,pdftex,moreauthors]{Definitions/mdpi} 

\firstpage{1} 
\pubvolume{1}
\issuenum{1}
\articlenumber{0}
\pubyear{2023}
\copyrightyear{2023}
\externaleditor{Academic Editor:{Fulai Guo}%please add academic editor 
}
\datereceived{11 May 2023} 
\daterevised{5 June 2023} % Comment out if no revised date
\dateaccepted{28 June 2023} 
\datepublished{ } 
\hreflink{https://doi.org/} % If needed use \linebreak
\Title{Investigating Possible Correlations between Gamma-Ray and Optical Lightcurves for TeV-Detected Northern Blazars over \mbox{8 Years} of Observations}

\TitleCitation{Investigating Possible Correlations between Gamma-Ray and Optical Lightcurves for TeV-Detected Northern Blazars over 8 Years of Observations}

\Author{Atreya Acharyya %Please carefully check the accuracy of names and affiliations. Changes will not be possible after proofreading.
 $^{1,*,\dagger}$\orcidA{} and Alberto~C.~Sadun $^{2,\dagger}$}

\AuthorNames{Atreya Acharyya Alberto~C.~Sadun}

\AuthorCitation{Acharyya, A.; Sadun, A.C.}
\address{%
$^{1}$ \quad Department of Physics and Astronomy, University of Alabama, Tuscaloosa, AL 35487, USA\\
$^{2}$ \quad Department of Physics, University of Colorado Denver, Campus Box 157, P.O.Box 173364 %MDPI: Please check whether the pobox is necessary and can be deleted, please make the affiliation format of the same country consistent.
,\linebreak Denver, CO 80217, USA; alberto.sadun@ucdenver.edu}

\corres{Correspondence %MDPI: We added the corresponding author according to the information on redmine, please check and confirm.
: aacharyya1@ua.edu}

\firstnote{These authors contributed equally to this work.}
\abstract{Blazars are a subclass of active galactic nuclei (AGN) having relativistic jets aligned within a few degrees of our line-of-sight and form the majority of the AGN detected in the TeV regime. The Fermi-Large Area Telescope (LAT) is a pair-conversion telescope, sensitive to photons having energies between 20 MeV and 2 TeV, and is capable of scanning the entire gamma-ray sky every three hours. Despite the remarkable success of the Fermi mission, many questions still remain unanswered, such as the site of gamma-ray production and the emission mechanisms involved. The Asteroid Terrestrial-impact Last Alert System (ATLAS) is a high cadence all sky survey system optimized to be efficient for finding potentially dangerous asteroids, as well as in tracking and searching for highly variable and transient sources, such as AGN. In this study, we investigate possible correlations between the \textit{Fermi}-LAT observations in the 100 MeV--300 GeV energy band and the ATLAS optical data in the R-band, centered at 679 nm, for a sample of 18 TeV-detected northern blazars over 8 years of observations between 2015 and 2022. Under the assumption that the optical and gamma-ray flares are produced by the same outburst propagating down the jet, the strong correlations found for some sources suggest a single-zone leptonic model of emission.}


\begin{document}

%%%%%%%%%%%%%%%%%%%%%%%%%%%%%%%%%%%%%%%%%%
\setcounter{section}{0} %% Remove this when starting to work on the template.
%\section{How to Use this Template}

% The order of the section titles is different for some journals. Please refer to the "Instructions for Authors” on the journal homepage.

\section{Introduction}

Blazars form a subclass of radio-loud active galactic nuclei (AGN) that have jets closely aligned to our line-of-sight, resulting in the emission from these objects being highly Doppler-boosted and making them some of the brightest gamma-ray sources in the extragalactic sky~\citep{gammaagn}. Blazars are generally characterized by non-thermal, highly-polarized continuum emission, spanning the entire electromagnetic spectrum and characteristically show very fast variability, which has been observed down to timescales of minutes in the gamma-ray regime~\citep{1996_Gaidos, PKS_2155_minute_variability, 2020_Mrk421}, {{as well as in the optical regime~\citep{2018_Aranzana, 2018_Kim, galaxies6010034}}.} 

The spectral energy distribution (SED) of a typical blazar comprises two distinct peaks. Although the first peak, occurring in the radio to the X-ray regime, has been attributed to synchrotron emission from electrons and positrons within the jet, the physical mechanisms responsible for the second peak, produced in the X-ray to gamma-ray regime, is still a matter of debate and two main scenarios have been postulated to explain it. Leptonic models \citep{Blandford_and_Levinson_1995, Georganopoulos_2002} attribute the high-energy peak to the inverse Compton (IC) scattering between the energetic leptons in the jet and a field of low energy photons (for example, \cite{1994_Sikora}), either the same photons emitted through synchrotron emission (synchrotron self-Compton (SSC)) or photon populations external to the jet (external inverse Compton (EIC)).
On the other hand, hadronic models (for example, \cite{MB_2013}) suggest that the second peak may be a result of high energy photons produced in  cosmic-ray interactions through the decay of neutral pions (for example, \cite{1992_Mannheim}) or proton synchrotron emission {{(for example, \cite{2016_kovalev}).}}

Localizing the gamma-ray emission is an indirect process and a variety of different methods have been used previously in the literature. 
{{Assuming constant jet geometry, the size of the emission region, r, has been used to infer its distance from the supermassive black hole (SMBH), R, using $\text{r}=\psi \text{R}$. Here $\psi$ is the semi-aperture opening angle of the jet \citep{Ghisellini_2009, 2009Dermer}. This relation has been used to constrain the emission region to be within a few parsec from the base of the jet. For example, Ref. \cite{Foschini_2011}, % we added ref. before ref citation, please check and modify.
 using $\sim$2 years of \textit{Fermi}-LAT observations, constrained the emission to be from within the broad line region (BLR) under the assumption that the full width of the jet is responsible for the emission.}}

{{Moreover, the observation of very high energy (VHE) photons ($E_{\gamma} \geq 20$ GeV) suggest the emission originates farther out, at parsec scale distances and from within the molecular torus (MT) region \citep{Donea_2003, RN28}. VHE photons are expected to be severely attenuated from interactions with the photons in the BLR and the detection of blazars with ground-based instruments is difficult to explain if the emission were assumed to originate in regions near the central engine. A possible solution which accommodates both the short variability timescales and VHE detection is to abandon the one-zone emission model and invoke the presence of multiple emission regions \cite{2021MNRAS.500.5297A}.}}

An important technique to localize the emission region in blazars is the correlation of multi-wavelength lightcurves. In particular, it helps in identifying potential relationships between emission zones and in understanding the dominant emission mechanisms. For instance, a strong correlation between synchrotron produced optical data and IC scattering produced gamma-ray observations {{(for example, \citep{2019_Liodakis, 2023_de_Jaeger})}} is expected in some leptonic models. The lags and leads, if highly significant, can help to constrain the location of the gamma-ray emission region relative to the {{optical emission region}} and also discriminate between SSC and EC models. On the other hand, orphan flares in one waveband having no correlation with others, for example, the 2002 flare of 1ES 1959+650 \citep{2004_orphan}, can be interpreted as evidence of multiple emission zones or as support for hadronic emission models. 

In this respect, the Large Area Telescope (LAT) on board the \textit{Fermi} satellite \cite{Fermi_LAT} has been particularly important. This pair-conversion telescope, launched in June 2008, is sensitive to photon energies between 20 MeV and 2 TeV and scans the entire gamma-ray sky every three hours. For example,  Ref. \cite{Cohen_2014}% we added ref. before ref citation, please check and modify.
 presented an investigation of correlations between optical data collected with the robotic 0.76 m Katzman Automatic Imaging Telescope (KAIT) at Lick Observatory and gamma-ray observations with the \textit{Fermi}-LAT using discrete correlation functions (DCFs). The same method was used by \cite{Fuhrmann,Max_Moerbeck_2014} to study radio–gamma-ray correlations in blazars.

{{The Asteroid Terrestrial-impact Last Alert System (ATLAS~\cite{ATLAS_main}) is a high cadence all sky survey system comprising four independent units, one on Haleakala (HKO), and one on Mauna Loa (MLO) in the Hawaiian islands in the Northern Hemisphere and one each at the El Sauce Observatory, Chile and the South African Astronomical Observatory in the Southern Hemisphere. It is optimized to be efficient for finding potentially dangerous asteroids, as well as in tracking and searching for highly variable and transient sources, such as AGN, and is capable of discovering more bright, less than 19 mag, supernovae candidates than other ground based surveys.}} 

Blazar observations with ATLAS are in the R-band, centered at 679~nm, having a typical cadence of one data point per two days. In this work, we apply local cross-correlation functions (LCCFs~\cite{Welsh_LCCF}) to investigate possible correlations between ATLAS optical data and gamma-ray observations with the \textit{Fermi}-LAT for a sample of 18 TeV-detected northern blazars, over 8 years of observations between 2015 and 2022. {{For brighter sources (for example, Mrk 501 \cite{galaxies6010020}), it is also possible to undertake an analysis of microvariability in outbursts using optical data and this will be investigated in detail in a future publication.}}

\section{Source Selection and Data Reduction}

The main goal of this study is to investigate possible correlations between ATLAS optical data and \textit{Fermi}-LAT gamma-ray observations in blazars. The identification of suitable sources was primarily governed by having sufficient photon statistics to allow for a detailed study of the LCCFs. The sample of sources chosen for this study comprised 18 \textit{Fermi}-LAT northern blazars detected in the TeV regime \footnote[1]{\url{http://tevcat2.uchicago.edu/} (accessed on 3 June  2023)} with ground-based gamma-ray observatories is shown in Table~\ref{Table 1.}.

Optical observations were made with the ATLAS, a high cadence all sky survey system of four 0.5~m telescopes that scan the entire sky periodically.  The telescopes are located in Hawaii (Mauna Loa, Hawaii and Haleakala, Maui), in South Africa (Sutherland Observing Station), and Chile (El Sauce Observatory, Rio Hurtado). Between declinations $-$50 degrees and + 50 degrees, the cadence is one set of observations each day, and in the polar regions once every two days, during observing season and weather permitting. The filter used in these observations is roughly equivalent to the Johnson--Cousins R filter.  The transmission curve is centered at 679 nm.  The filter is nonstandard, but well calibrated, see \cite{ATLAS_main}.  All the data is archived in a database for retrieval.

\begin{figure}[H]
\centering
\includegraphics[width= 0.452 \textwidth]{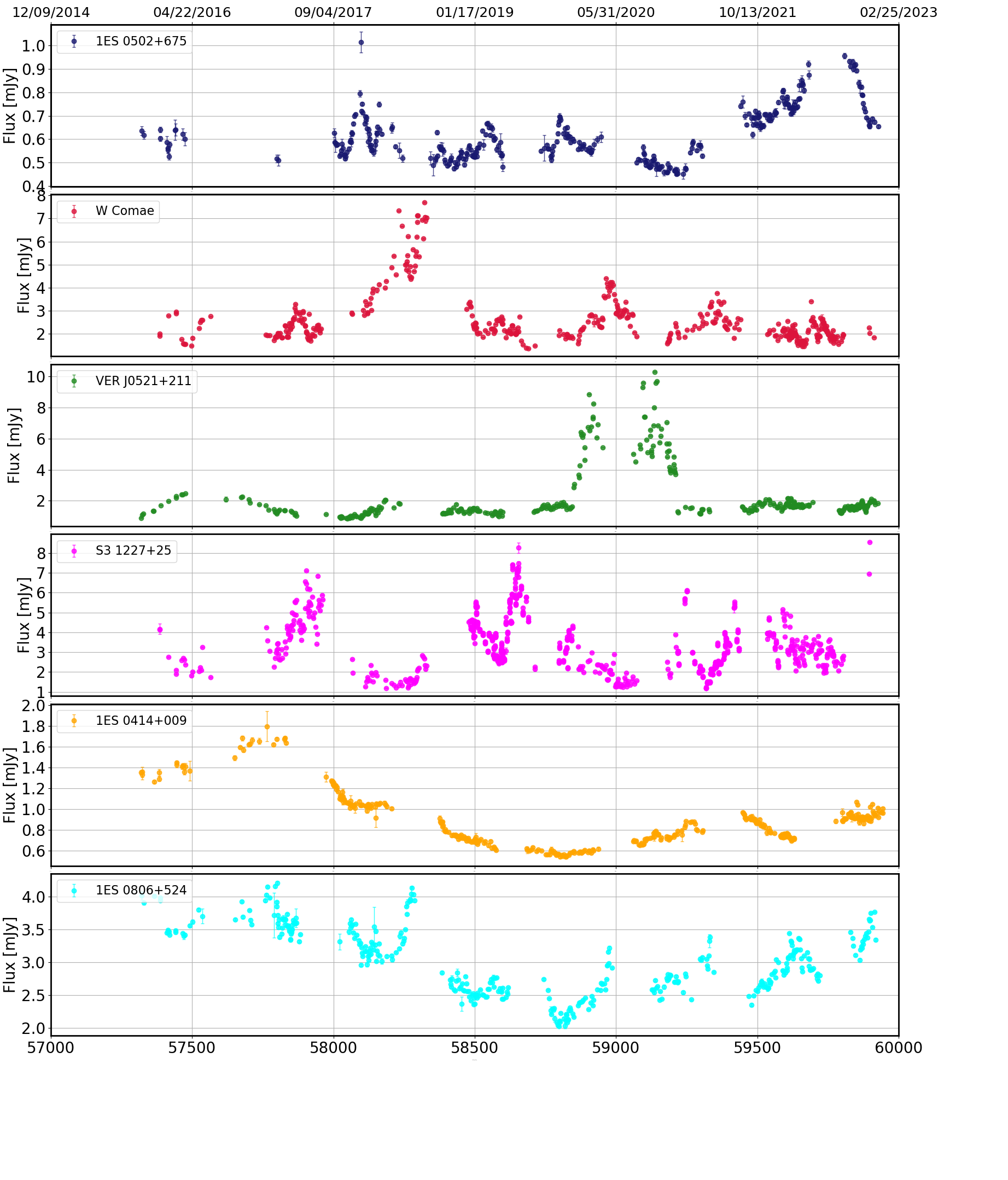}
\includegraphics[width= 0.49  \textwidth]{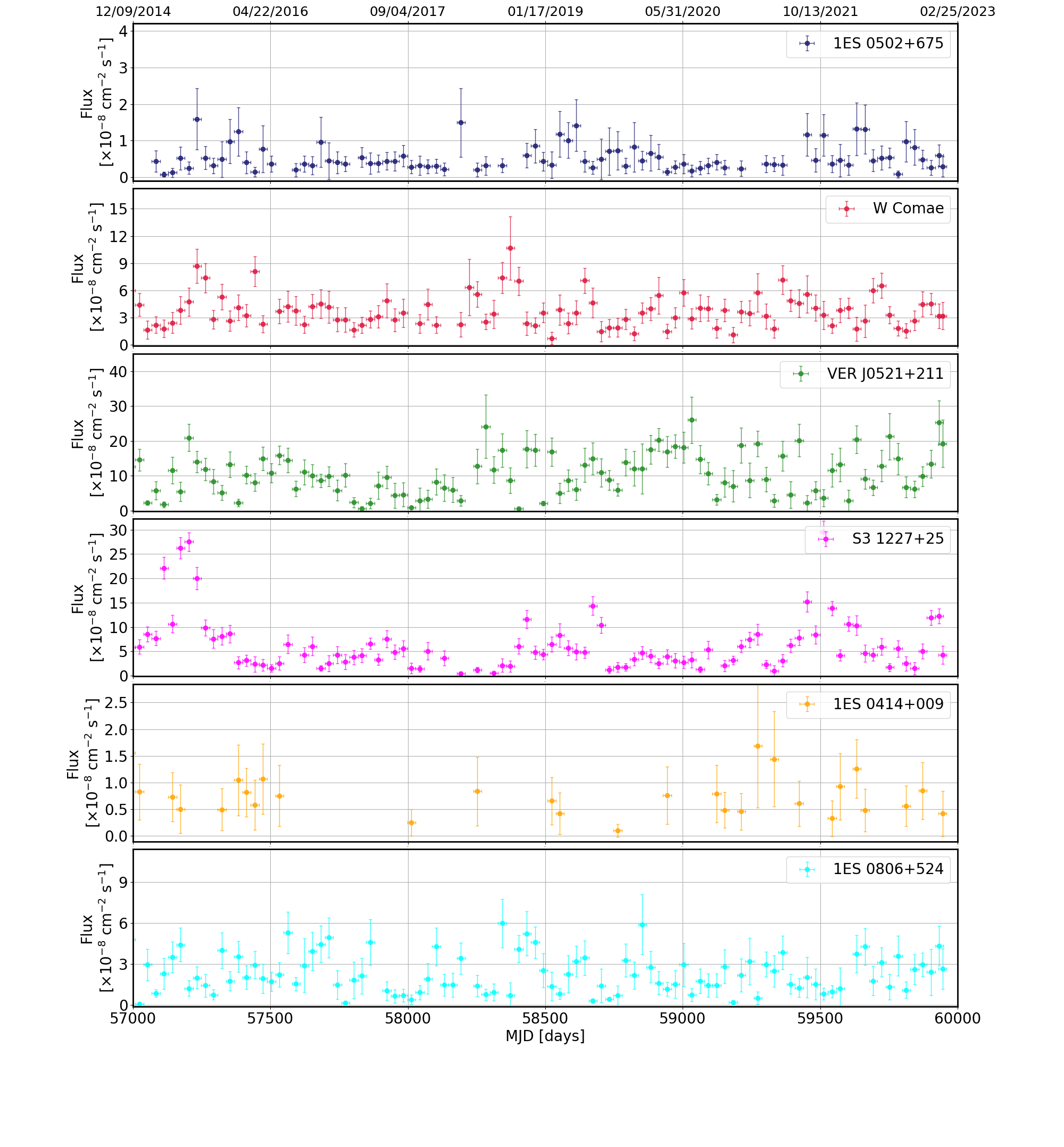}
\vspace{-15pt}
\caption {\textbf{Left}: Optical flux observations for 1ES 0502+675, W Comae, VER J0521+211, S3 1227+25, 1ES 0414+009, and 1ES 0806+524 between 9 December 2014 (MJD {57000} 
) and 1 January~2023 (MJD {59945}) obtained with ATLAS in the R band data. The errors are purely statistical. \textbf{Right}: The corresponding gamma-ray lightcurves for 1ES 0502+675, W Comae, VER J0521+211, S3 1227+25, 1ES 0414+009, and 1ES 0806+524 between 9 December 2014 (MJD {57000}) and 1 January 2023 (MJD {59945}) binned in monthly periods. The errors are purely statistical and only data points with TS $\geq$ 10 are shown.}
\label{fig: Figure 1.}
\end{figure}

\begin{figure}[H]
\centering
\includegraphics[width= 0.455 \textwidth]{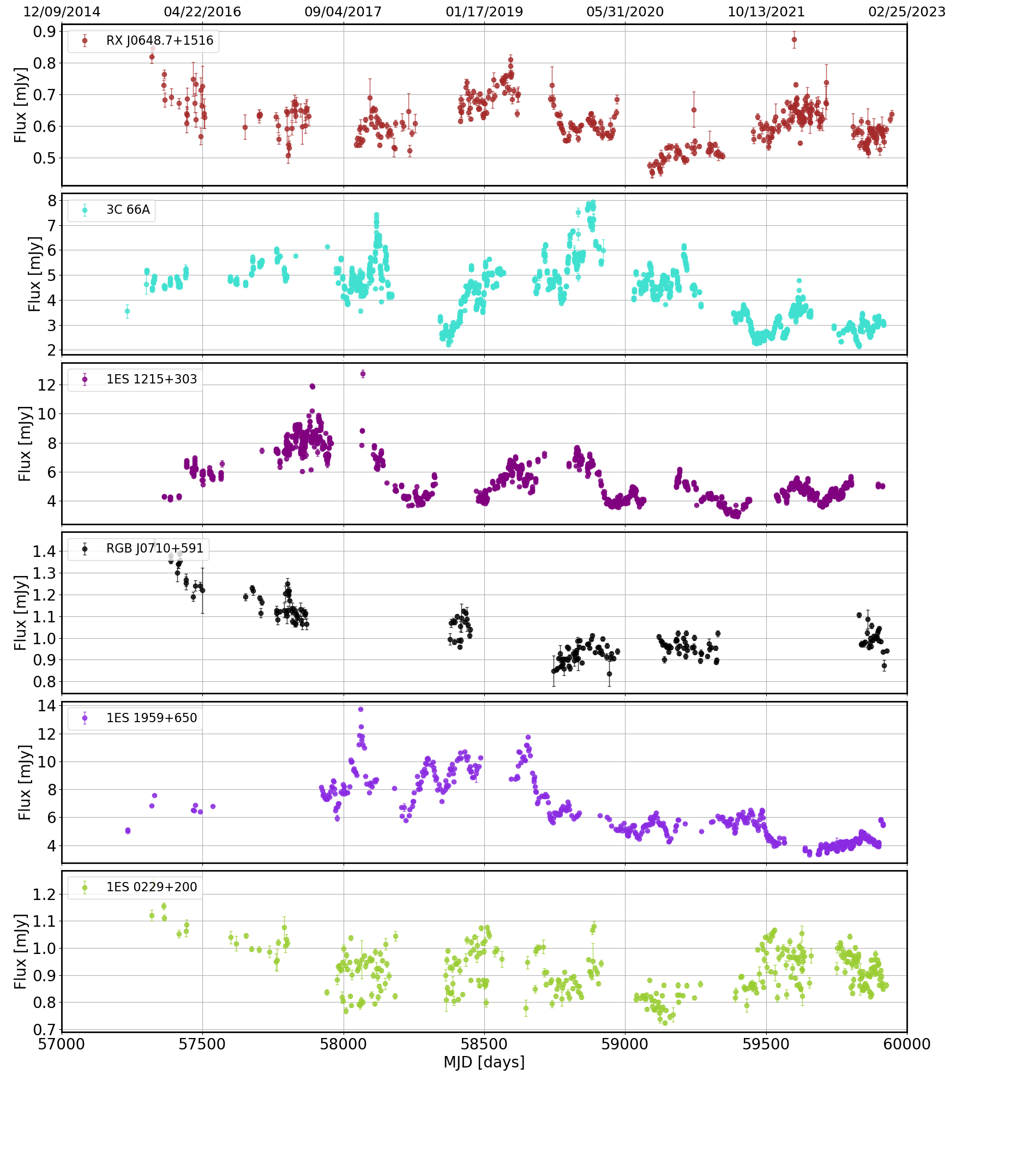}
\includegraphics[width= 0.49 \textwidth]{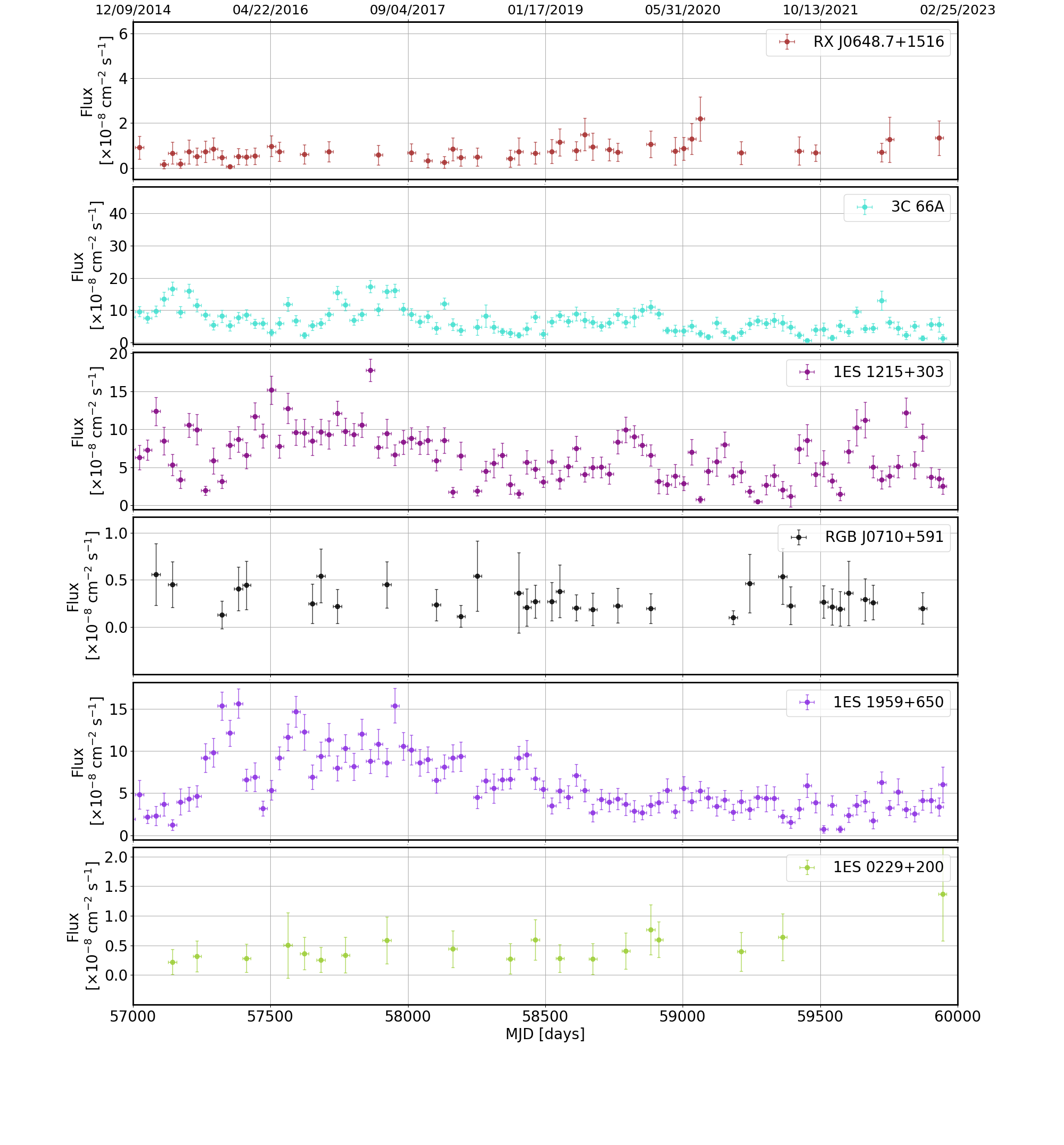}
\vspace{-15pt}
\caption {\textbf{Left}: Optical flux observations for RX J0648.7+1516, 3C 66A, 1ES 1215+303, RGB J0710+591, 1ES 1959+650, and 1ES 0229+200 between 9 December  2014 (MJD {57000}) and 1 January~2023 (MJD {59945}) obtained with ATLAS in the R band data. The errors are purely statistical. \textbf{Right}: The corresponding gamma-ray lightcurves for RX J0648.7+1516, 3C 66A, 1ES 1215+303, RGB J0710+591, 1ES 1959+650, and 1ES 0229+200 between 9 December  2014 (MJD {57000}) and 1 January~2023 (MJD {59945}) binned in monthly periods. The errors are purely statistical and only data points with TS $\geq$ 10 are shown.}
\label{fig: Figure 2.}
\end{figure}

\begin{figure} [H]
\centering
\includegraphics[width= 0.453 \textwidth]{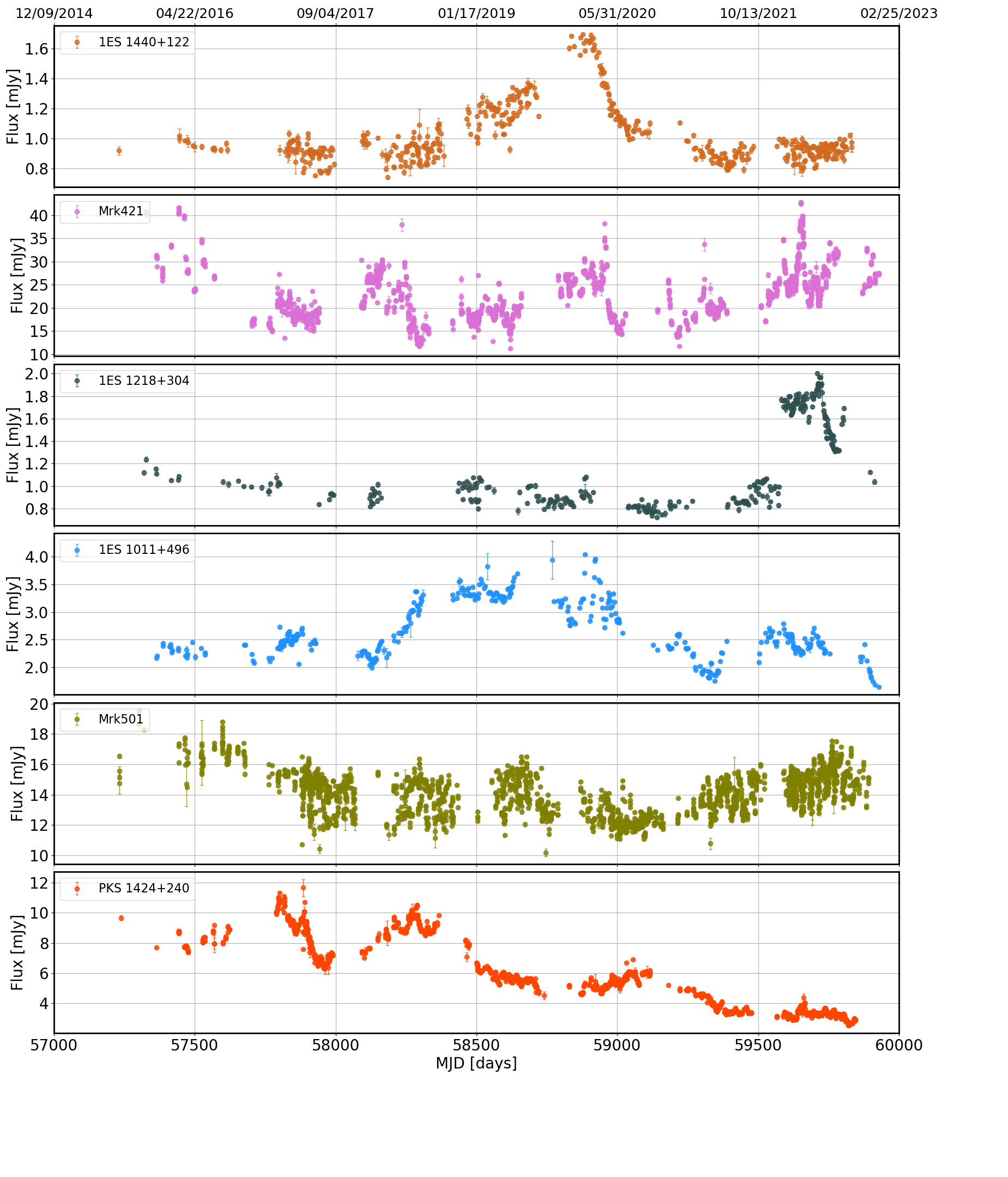}
\includegraphics[width= 0.49 \textwidth]{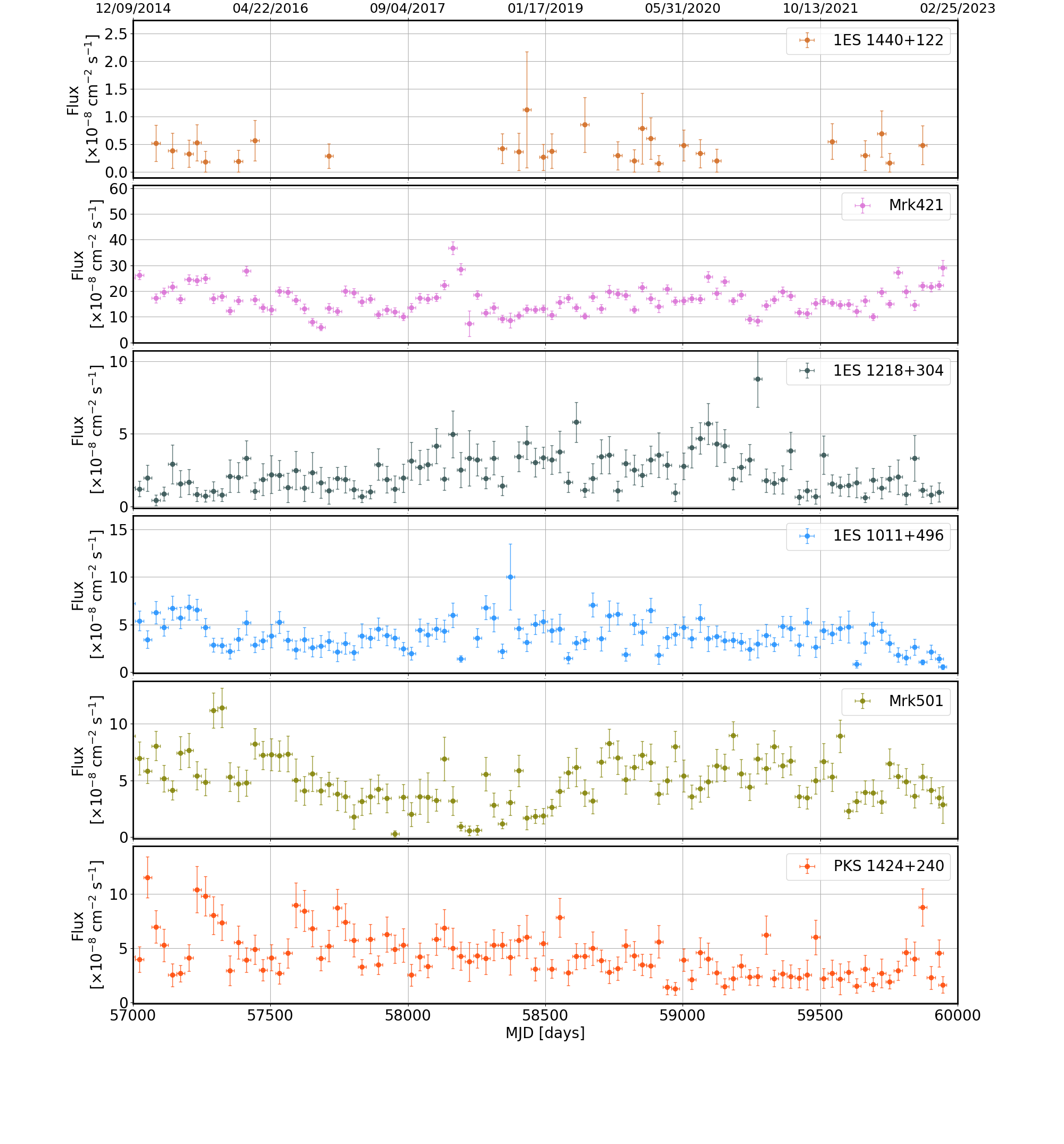}
\vspace{-15pt}
\caption {\textbf{Left}: Optical flux observations for 1ES 1440+122, Mrk 421, 1ES 1218+304, 1ES 1011+496, Mrk 501, and PKS 1424+240 between 9 December 2014 (MJD {57000}) and 1 January 2023 (MJD {59945}) obtained with ATLAS in the R band data. The errors are purely statistical. \textbf{Right}: The corresponding gamma-ray lightcurves for 1ES 1440+122, Mrk 421, 1ES 1218+304, 1ES 1011+496, Mrk 501, and PKS 1424+240 between 9 December 2014 (MJD {57000}) and 1 January 2023 (MJD {59945}) binned in monthly periods. The errors are purely statistical and only data points with TS $\geq$ 10 are shown.}
\label{fig: Figure 3.}
\end{figure}

A forced photometry system is used in which {{the point spread function (PSF), which represents the distribution of light from a point source on a detector}}, is obtained for each object based on bright stars, and the function is forced onto the object at the chosen coordinates.  Data processing is described in \cite{ATLAS_main,ATLAS_smith}.  When the data is retrieved from the database, software automatically calculates the magnitude according to the AB system, as well as the flux for each object.  A catalog of variable stars produced by this method is shown by \cite{2018_Heinze}. The optical lightcurves {{for the sample of blazars}} are shown in Figures \ref{fig: Figure 1.}--\ref{fig: Figure 3.}.% we changed the position of the Figures 1-5 to make them after the first mention, please confirm.

In this work, we analyzed \textit{Fermi}-LAT photons detected between MJD~{57000} and MJD~{59945}, which corresponds to midnight on 9 December 2014 until midnight on\linebreak \mbox{1 January~2023.} Throughout the gamma-ray analysis, we used the \textit{Fermi} Science Tools version 11-05-03~\footnote[2]{\url{http://fermi.gsfc.nasa.gov/ssc/data/analysis/software} (accessed on 3 June 2023)}, \textit{FERMIPY} version 1.0.1~\footnote[3]{\url{http://fermipy.readthedocs.io} (accessed on 3 June 2023)} \citep{wood2017fermipy}, in conjunction with the latest \textit{PASS} 8 IRFs~\citep{atwood2013pass}. We consider the energy range 100 MeV--300 GeV and a region of interest (RoI) with radius 15$^{\circ}$ centred on each source. 
A lower limit of 100 MeV is chosen because the PSF of the \textit{Fermi}-LAT increases at lower energies, making a point source analysis difficult. Furthermore, most of the blazars selected have relatively soft gamma-ray spectra and are not expected to be significantly detected with the \textit{Fermi}-LAT at energies above 300 GeV.
Moreover, we selected only photon events from within a maximum zenith angle of 90$^{\circ}$ in order to reduce contamination from background photons from the Earth's limb, produced by the interaction of cosmic-rays with the upper atmosphere.

Sources in the 4FGL-DR3 catalog \cite{4fgl_dr3} within a radius of $20^{\circ}$ from the spatial position of each source in the sample were included in the initial model for the analyses, with their spectral parameters fixed to their catalog values.
This takes into account the gamma-ray emission from sources lying outside the RoI which might yet contribute photons to the data, especially at low energies, due to the size of the point spread function of the \textit{Fermi}-LAT. The contributions from the isotropic and galactic diffuse backgrounds were modeled using the most recent templates, iso\_P8R3\_SOURCE\_V2\_v1.txt and gll\_iem\_v07.fits, respectively. 

The analysis began with an initial automatic optimization of the RoI by iteratively fitting the sources. This ensured all parameters were close to their global likelihood maxima. The spectral normalization of all modelled sources within the RoI were left free as were the normalization factor of both the isotropic and galactic diffuse emission templates. Furthermore, the spectral shape parameters of all sources within 5$^{\circ}$ of the centre of the RoI were left free to vary whereas those of other sources were fixed to the values reported in the 4FGL-DR3 catalog. A binned likelihood analysis was then performed in order to obtain the spectral parameters best describing the model using a spatial binning of 0.1$^{\circ}$ pixel$^{-1}$ and \mbox{8 energy} bins per decade.

In order to pursue a study of the temporal behaviour of the gamma-ray fluxes, the \mbox{15 year} \textit{Fermi}-LAT data were binned monthly with a likelihood routine applied to each bin separately \footnote[4]{This was implemented using the \textit{gta.lightcurve()} %MDPI: Is italics necessary? Can it be removed.
 method in \textit{FERMIPY}}.. During the production of the lightcurve, the spectral parameters of all sources within 5$^{\circ}$ of the RoI centre were left free for each bin as were the normalization factors of the background emission models. The resulting lightcurves {{for the sample of blazars}} are shown in  Figures \ref{fig: Figure 1.}--\ref{fig: Figure 3.}, along with the corresponding uncertainties. Only time intervals having  TS $\geq$ 10 were considered, which roughly equates to a significance of $3\sigma$. 

\section{Results}

Local cross-correlation functions (LCCFs; \cite{Welsh_LCCF}) were then applied to investigate correlations between the \textit{Fermi}-LAT and ATLAS lightcurves. {{Consider two lightcurves having fluxes $a_{i}$ and $b_{j}$, corresponding to times $t_{a_{i}}$ and $t_{b_{j}}$, the LCCF can then be computed as:}}

{{
\begin{equation}
    \centering
    \text{LCCF} (\tau) = \frac{1}{M}\frac{\Sigma (a_{i} - \overline{a}_{\tau}) (b_{j} - \overline{b}_{\tau})}{\sigma_{a \tau}\sigma_{b \tau}}.
    \label{eq:LCCF}
\end{equation}
}}

\begin{figure} [H]

\includegraphics[width= \textwidth]{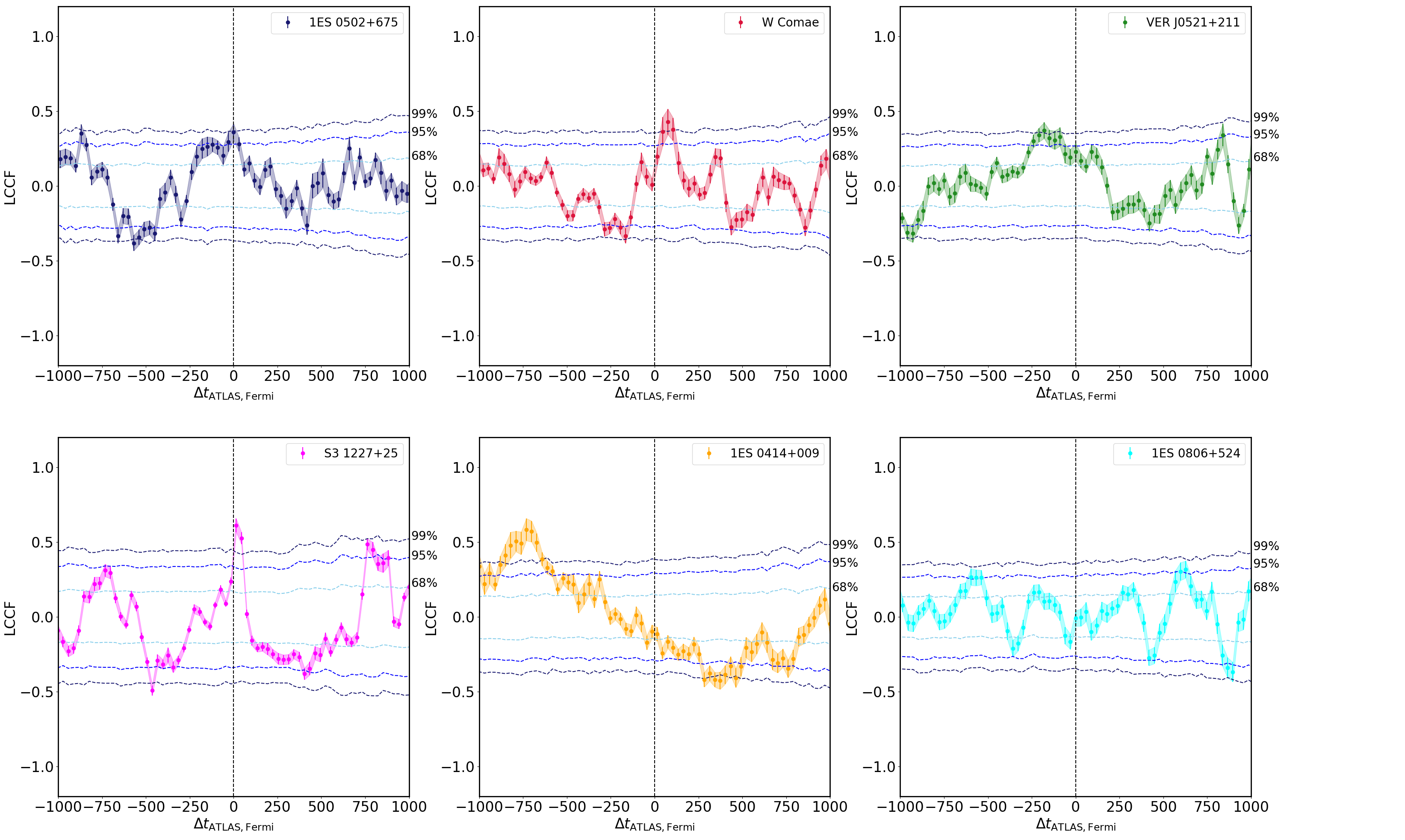}
\caption {Local cross-correlation functions (LCCFs) calculated between the \textit{Fermi}-LAT and ATLAS lightcurves for 1ES 0502+675, W Comae, VER J0521+211, S3 1227+25, 1ES 0414+009, and 1ES 0806+524. The x-axis represents the time delay, $\Delta t _{\text{ATLAS},\text{Fermi}} = t_{\text{ATLAS}} -t_{\text{Fermi}}$, in days for two lightcurves ATLAS and \textit{Fermi}-LAT. The dashed  vertical line corresponds to $\Delta t _{\text{ATLAS},\text{Fermi}} = 0$. The corresponding shaded regions indicate the error bounds of the LCCFs. The blue lines represent the 68$\%$, 95$\%$, and 99$\%$ confidence intervals (from lighter to darker shades) derived from Monte Carlo simulations.}
\label{fig: Figure 4.}
\end{figure}

\begin{figure}[H]

\includegraphics[width=  \textwidth]{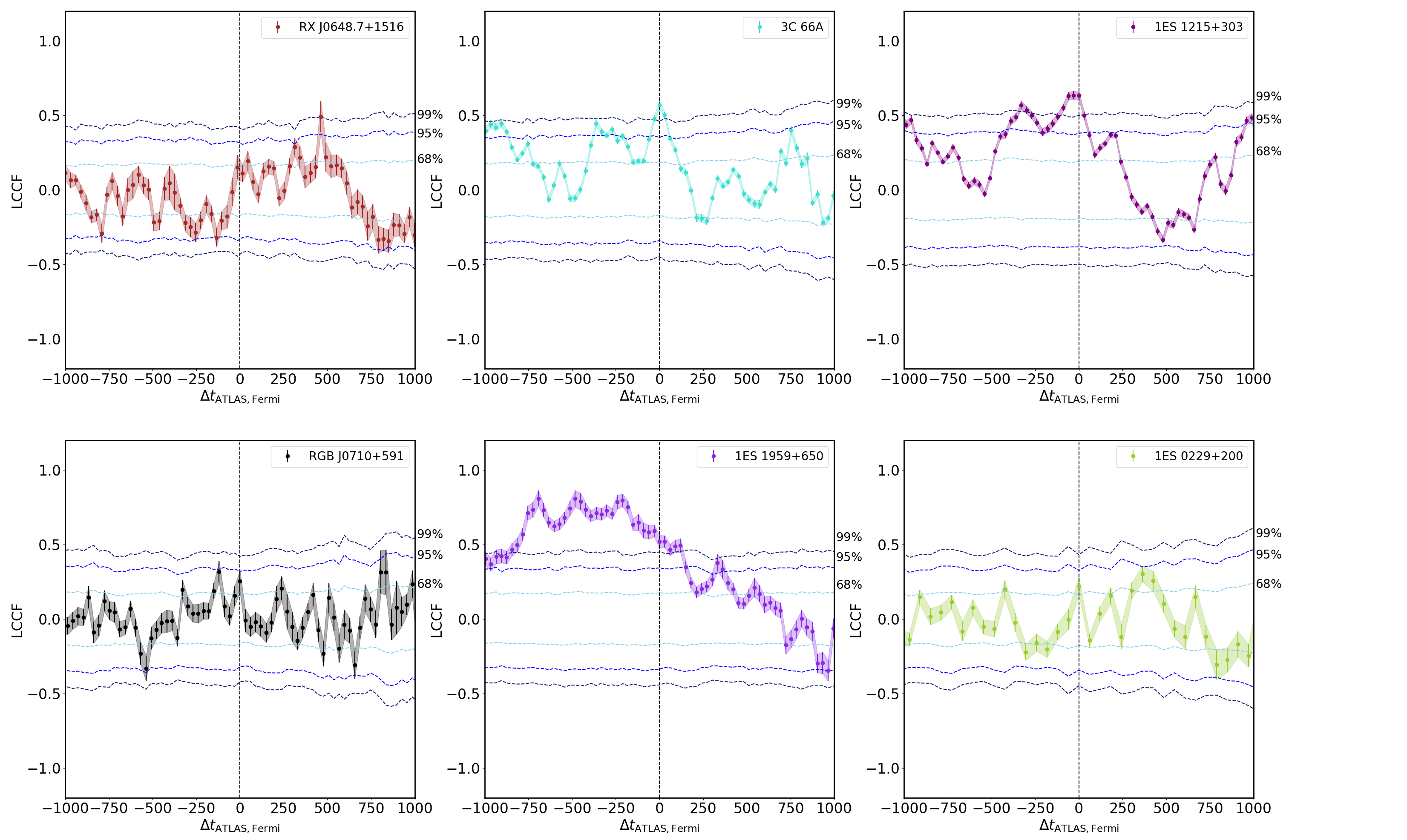}

\includegraphics[width=  \textwidth]{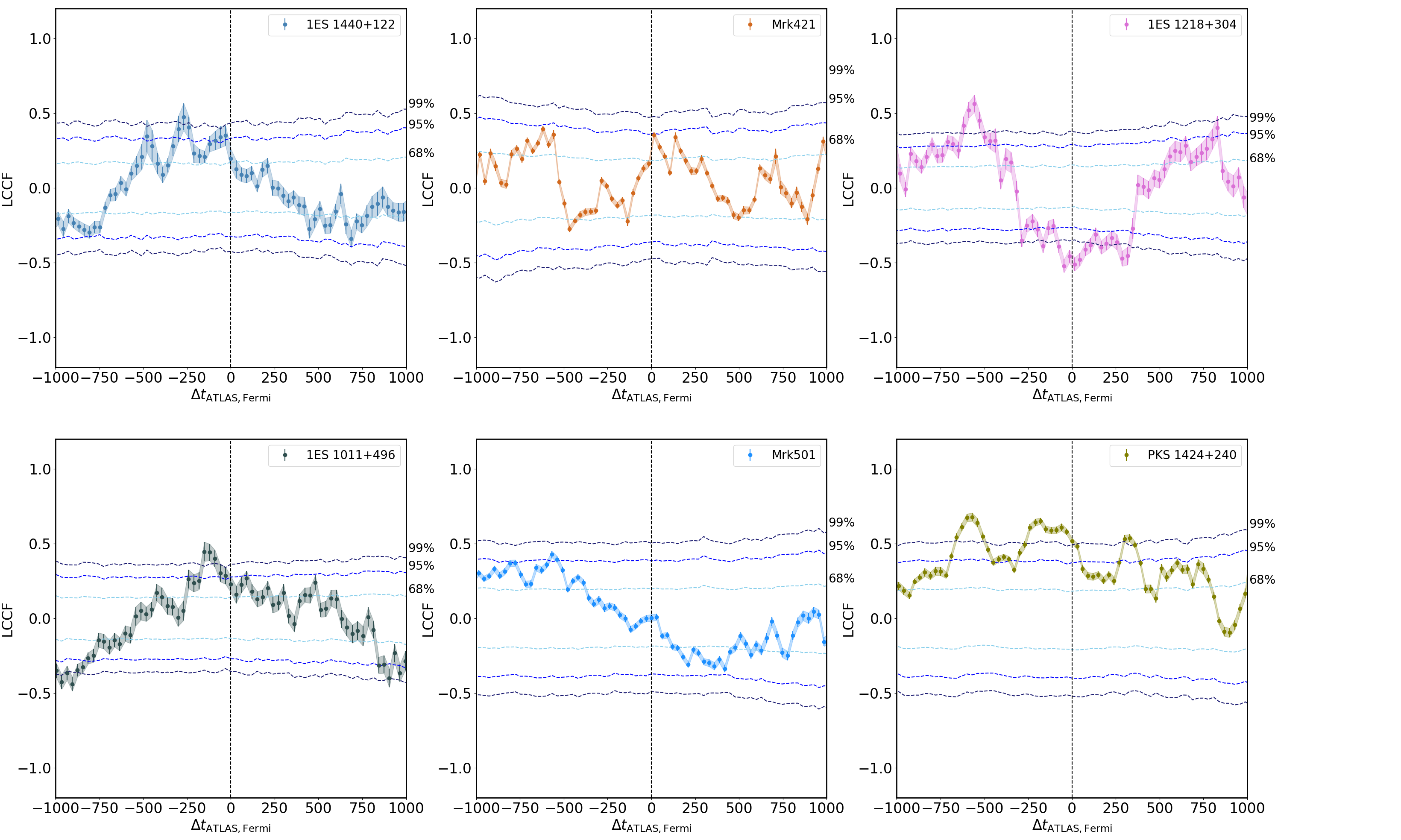}
\caption {\textbf{Top}: Local cross-correlation functions (LCCFs) calculated between the \textit{Fermi}-LAT and ATLAS lightcurves for RX J0648.7+1516, 3C 66A, 1ES 1215+303, RGB J0710+591, 1ES 1959+650, and 1ES 0229+200. The x-axis represents the time delay, $\Delta t _{\text{ATLAS},\text{Fermi}} = t_{\text{ATLAS}} -t_{\text{Fermi}}$, in days for two lightcurves ATLAS and \textit{Fermi}-LAT. The dashed  vertical line corresponds to $\Delta t _{\text{ATLAS},\text{Fermi}} = 0$. The corresponding shaded regions indicate the error bounds of the LCCFs. The blue lines represent the 68$\%$, 95$\%$, and 99$\%$ confidence intervals (from lighter to darker shades) derived from Monte Carlo simulations. \textbf{Bottom}: Local cross-correlation functions (LCCFs) calculated between the \textit{Fermi}-LAT and ATLAS lightcurves for 1ES 1440+122, Mrk 421, 1ES 1218+304, 1ES 1011+496, Mrk 501, and PKS 1424+240. The x-axis represents the time delay, $\Delta t _{\text{ATLAS},\text{Fermi}} = t_{\text{ATLAS}} -t_{\text{Fermi}}$, in days for two lightcurves ATLAS and \textit{Fermi}-LAT. The dashed  vertical line corresponds to $\Delta t _{\text{ATLAS},\text{Fermi}} = 0$. The corresponding shaded regions indicate the error bounds of the LCCFs. The blue lines represent the 68$\%$, 95$\%$, and 99$\%$ confidence intervals (from lighter to darker shades) derived from Monte Carlo simulations.}
\label{fig: Figure 5.}
\end{figure}

{{Here, the sums run over $M$ pairs for which $\tau \leq t_{a_{i}}-t_{b_{j}}<\tau+\Delta t$ for a chosen timestep $\Delta t$, and $\overline{a}_{\tau}$ and $\overline{b}_{\tau}$ are flux averages and $\sigma_{a \tau}$ and $\sigma_{b \tau}$ are standard deviations over the $M$%Please unify the format of the variable in the full text.
~pairs, respectively \citep{Welsh_LCCF}. As adopted in \cite{2019_Meyer}, we choose the higher half median of the time separations between consecutive data points in the two lightcurves as the binning of the timelags, $\tau$. Furthermore, the minimum and maximum values of $\tau$ are chosen to be $\pm 0.5$ times the length of the shorter lightcurve \citep{Max_Moerbeck_CCFs}.}}
This method is independent of any difference in sampling rates between the lightcurves. LCCFs are intrinsically bound in the interval [$-$1,1] and \citep{Max_Moerbeck_CCFs} found them to be more efficient than the use of Discrete Correlation Functions (DCFs; \cite{RN19}). The LCCFs obtained for each source are shown in Figures \ref{fig: Figure 4.} and \ref{fig: Figure 5.}.

As done in \cite{Cohen_2014}, the centroid lags and uncertainties are derived from weighted least-square Gaussian fits to the LCCF points at the {{location of the most significant correlation peak}}. Although the peaks of the Gaussian fits give a first order determination of the uncertainty, these do not account for the effects of correlated red-noise between the datasets \cite{Uttley}. The significances of the correlations are obtained by performing Monte Carlo simulations to produce 1000 artificial lightcurves matching the probability distribution function (PDF) and power spectral density (PSD) of each observation using the method outlined in \cite{Emmanoulopoulos} \footnote[5]{The code was developed from Connolly, S. D., 2016, Astrophysics Source Code Library, record ascl:1602.012. See https://github.com/samconnolly/DELightcurveSimulation, (accessed on 3 June 2023) %MDPI: please add accesse date. (day month year, e.g 25 June 2023).
.}%we changed footnote 6 into \footnote 5, please check and confirm
.  The 68$\%$, 95$\%$, and 99$\%$ confidence intervals obtained are also shown in Figure \ref{fig: Figure 4.}, in blue from lighter to darker shades.

The centroid lags and significances obtained for each source are summarised in Table~\ref{Table 1.}. Throughout this paper, a positive time delay, $\Delta t > 0$, is defined as corresponding to the optical emission leading the gamma-ray emission. Wider LCCF peaks may indicate a range of characteristic timescales in the correlated response or limitations in the instruments~\cite{Cohen_2014}. The presence of a peak, significant at a level of above 3$\sigma$, implies strong optical and gamma-ray correlations. This is seen for W Comae, S3 1227+25, 3C 66A, and 1ES 1215+303, and suggests a single-zone model of emission. Under the assumption that the optical and gamma-ray flares are produced by the same outburst propagating down the jet, both positive and negative time-lags on timescales of days to {{tens of days}} are predicted in SSC and EIC models. A significant peak consistent with zero indicates an absence of time-lag and is seen for S3 1227+25 and 3C 66A.

{{The results obtained in this investigation are broadly in agreement with those seen in \cite{2019_Liodakis}, where the optical and gamma-ray correlations of 121 blazars were found to be significant at a level of above $68 \%$ ($\sim 1 \sigma$). The majority of the corresponding time-lags were within $\pm 20$ days of zero-lag and this was interpreted as evidence of a common origin for the flares in the two bands, implying leptonic processes dominate blazar gamma-rays, while not excluding the possibility of hadronic contribution to the emission. However, it should be noted that although \cite{2019_Liodakis} found that the time delays, in general, did not exceed \mbox{50 days} at $2 \sigma$ level and was less at $3 \sigma$ level, in this study, we find significant delays for 11 of the 18 considered objects, often exceeding 100 days.}}

{{Furthermore, out of the nine objects with a correlation of above $ 99\%$, five sources, namely, 1ES 0414+009, 1ES 1959+650, 1ES 1440+122, 1ES 1011+496, 1ES 1218+304, and PKS 1424+240, are found have a delay of more than 100 days. In all of these cases a negative time-lag is found, suggesting that the gamma-ray emission lags the optical emission. This long delay could imply a more complex emission mechanism than the simple one-zone leptonic SSC model found for the remaining sources; for example, one with multiple emission regions with the optical emission originating upstream of the gamma-ray emission (this was also seen for gamma-ray and radio correlations in \cite{Max_Moerbeck_2014}). However, continuous and simultaneous monitoring over a longer time period is required to draw any stronger conclusions.}}

{{Moreover, while similar conclusions of leptonic single-zone models where the low- and high-energy emission comes from the same population of electrons were also found in \cite{2023_de_Jaeger} from a study of gamma-ray and optical correlations for a sample of 1180 blazars, it should also be noted that they find evidence of "orphan" flares, gamma-ray flares with no optical counterpart or optical flares having no gamma-ray counterpart, in some of the sources.}} As seen in Figures \ref{fig: Figure 1.}--\ref{fig: Figure 3.}, a few sources in this investigation, {{for example, 1ES 0502+675 (MJD$\sim${59750}) and S3 1227+25 (MJD $\sim${58600})}}, also show "orphan" flares. The origin of these flares is uncertain but they have been interpreted as support for hadronic emission models (for example, \cite{2007Ap&SS.309...95B}), evidence of multiple emission zones (for example,~\cite{2021MNRAS.500.5297A}) or a result of contamination in the optical regime due to accretion disk emission (for example, \cite{2014ApJ...786..157A}).

\begin{table}[H]

\caption{The final~sample of \textit{Fermi}-LAT northern blazars detected in the TeV regime, selected for this study along with their classification, right ascensions (RA), and declinations (DEC) in degrees, redshifts ($z$; shown as "--" for sources not having a verified redshift measurement). The final two columns list the times corresponding to the peaks of the Gaussian fit along with the associated uncertainties and their significance in percentile derived from Monte Carlo simulations. Listed below are the TeV detection references for each source in the sample.}
\newcolumntype{C}{>{\centering\arraybackslash}X}
\begin{tabularx}{\textwidth}{m{2.5cm}<{\centering}m{1cm}<{\centering}m{1cm}<{\centering}m{1cm}<{\centering}m{1cm}<{\centering}m{2.5cm}<{\centering}m{2cm}<{\centering}}

\toprule
\textbf{Source} \vspace{-9pt}& \textbf{Class}	\vspace{-9pt}& \textbf{RA} & \textbf{Dec} & \pmb{\textbf{$z$}} \vspace{-9pt} & \textbf{Timelag} &\textbf{Significance}\\
 &   &\textbf{[deg]} &\textbf{[deg]} & &\textbf{[days]} &\textbf{[\%]}  \\
\midrule
1ES~0502+675~$^{[a]}$ &HBL	&76.9	&67.6 &0.340 &$-$9.75 $\pm$ 3.52  &$\geq$68\\ %& \\
W~Comae~$^{[b]}$	 &IBL	&185.4	&28.2 &0.102 &74.78 $\pm$ 6.85 &$\geq$99\\ %& \\
VER~J0521+211~$^{[c]}$  &IBL	&80.4	&21.2 &-- &$-$236.07 $\pm$ 6.65 &$\geq$95\\  %&\\
S3~1227+25~$^{[d]}$  &IBL &187.6	&25.3 &-- &20.71 $\pm$ 7.10 &$\geq$99\\ %&\\
1ES~0414+009~$^{[e]}$  &HBL	&64.2	&1.1 &0.287 &$-$746.48 $\pm$ 5.34 &$\geq$99\\ %& \\
1ES~0806+524~$^{[f]}$  &HBL	&122.5	&52.3 &0.138 &626.02 $\pm$ 6.65 &$\geq$95 \\ %&V \\
RXJ0648.7+1516~$^{[g]}$  &HBL &102.2	&15.3 &0.179 &466.59 $\pm$ 3.85 &$\geq$95\\ %&\\
3C~66A~$^{[h]}$ 	&IBL	&35.7 &43.0 &0.340 &2.61 $\pm$ 2.67 &$\geq$99 \\%&\\
1ES~1215+303~$^{[i]}$  &HBL	&184.5	&30.1 &0.131 &$-$56.10 $\pm$ 5.97 &$\geq$99\\ %&\\
RGB~J0710+591~$^{[j]}$ &EHBL	&107.6	&59.1 &0.125 &$-$130.00 $\pm$ 28.90 &$\geq$68\\ %& \\
1ES~1959+650~$^{[k]}$ &HBL	&299.9	&65.2 &0.048 &$-$683.44 $\pm$ 9.26 &$\geq$99\\ %& \\
1ES~0229+200~$^{[l]}$ &EHBL	&38.2	&20.3 &0.139 &331.57 $\pm$ 19.98 &$\geq$68\\ %& \\
1ES~1440+122~$^{[m]}$ &HBL	&220.7	&12.0 &-- &$-$272.09 $\pm$ 2.03 &$\geq$99\\ %&V \\
Mrk~421~$^{[n]}$	 &HBL	&166.1	&38.2 &0.031 &28.68 $\pm$ 7.08 &$\geq$68 \\ %& \\
1ES~1218+304~$^{[o]}$ &EHBL	&185.3	&30.2 &0.182 &$-$565.78 $\pm$ 2.09 &$\geq$99 \\ %& \\
1ES~1011+496~$^{[p]}$ &HBL	&153.8	&49.4 &0.212 &$-$122.25 $\pm$ 8.26 &$\geq$99\\ %&\\
Mrk~501~$^{[q]}$	 &HBL	&253.5	&39.8 &0.034 &$-$574.50 $\pm$ 9.65  &$\geq$68\\ %&\\ 
PKS~1424+240~$^{[r]}$ &HBL	&216.8	&23.8 &-- &$-$581.28 $\pm$ 1.78 &$\geq$99 \\ %& \\
\bottomrule
\end{tabularx}
\footnotesize{ TeV detection references: [$a$]%Please check whether these letters need to be uniformly changed to italic superscript format.~
VERITAS~\cite{1ES_0502+675}; [$b$] VERITAS~\cite{W_Comae_detection}; [$c$] VERITAS~\cite{VER_J0521+211_detection}; [$d$] VERITAS~\cite{S31227_discovery_paper}; [$e$] H.E.S.S.~\cite{1ES_0414+009_detection}; [$f$] VERITAS~\cite{1ES_0806+524_detection}; [$g$] VERITAS~\cite{1FGL_J0648.8+1516}; [$h$] VERITAS~\cite{3C66A_detection}; [$i$] MAGIC~\cite{1ES_1215+303_discovery}; [$j$] VERITAS~\cite{RGB_J0710+591_discovery}; [$k$] Utah~\cite{1ES_1959+650_discovery}; [$l$] H.E.S.S.~\cite{1ES_0229+200_discovery}; [$m$] VERITAS~\cite{1ES_1440+122_discovery}; [$n$] Whipple~\cite{Mrk421_discovery}; [$o$] MAGIC~\cite{1ES_1218+30_detection}; [$p$] MAGIC~\cite{1ES_1011+496_detection}; [$q$] Whipple~\cite{Mrk_501_discovery} and [$r$] MAGIC~\cite{PKS_1424+240_detection}.}

\label{Table 1.}
\end{table}

\section{Conclusions}

In this work, we present an investigation into possible correlations between ATLAS optical data and gamma-ray observations with the \textit{Fermi}-LAT for a sample of 18 TeV-detected northern blazars, over 8 years of observations between 2015 and 2022. Overall, we find the optical and gamma-ray emission to be highly correlated in our sample, with varied  time delays, ranging from timescales of days to even years for some sources. With the exception of one source, 1ES~1218+304, all the correlations are found to correspond to a significance of $68 \%$ ($\sim 1 \sigma$) and for nine sources the correlations are found to correspond to significance of $99 \%$ ($\sim 3 \sigma$). 

The observed strong correlations support leptonic models of IC scattering gamma-ray emission in which the seed photons are scattered to higher energy in the relativistic jet by the same electrons responsible for synchrotron emission. It should be noted that the significance of the correlations and the corresponding time delays do not allow us to make strong conclusions on whether the seed photons are dominated by SSC or EIC radiation. However, the lack of clear trend towards lag or lead in our BL Lac sample agrees with the results presented in \cite{Cohen_2014} and can be interpreted as evidence towards SSC being the dominant mechanism in BL Lac sources, as opposed to FSRQs which, in general, show the gamma-ray leading the optical emission, interpreted as evidence towards EIC emission being the dominant mechanism.

In conclusion, the gamma-ray optical correlation in BL Lac sources appears complex, as also seen in other variability studies (for example \cite{Cohen_2014, 2023_de_Jaeger}). Multi-wavelength studies at high cadence over many years is needed to probe the emission mechanisms further and this will be possible with further coverage with ATLAS and other optical instruments in conjunction with the continued successful operation of the \textit{Fermi}-LAT. Finally, we aim to perform a comprehensive study of the sample further investigating micro-variability in both energy bands in the future.

%%%%%%%%%%%%%%%%%%%%%%%%%%%%%%%%%%%%%%%%%%
\vspace{6pt} 

%%%%%%%%%%%%%%%%%%%%%%%%%%%%%%%%%%%%%%%%%%
%% optional
%\supplementary{The following supporting information can be downloaded at:  \linksupplementary{s1}, Figure S1: title; Table S1: title; Video S1: title.}


%%%%%%%%%%%%%%%%%%%%%%%%%%%%%%%%%%%%%%%%%%

\authorcontributions{%For research articles with several authors, a short paragraph specifying their individual contributions must be provided. The following statements should be used ``Conceptualization, X.X. and Y.Y.; methodology, X.X.; software, X.X.; validation, X.X., Y.Y. and Z.Z.; formal analysis, X.X.; investigation, X.X.; resources, X.X.; data curation, X.X.; writing---original draft preparation, X.X.; writing---review and editing, X.X.; visualization, X.X.; supervision, X.X.; project administration, X.X.; funding acquisition, Y.Y. All authors have read and agreed to the published version of the manuscript.'', please turn to the  \href{http://img.mdpi.org/data/contributor-role-instruction.pdf}{CRediT taxonomy} for the term explanation. Authorship must be limited to those who have contributed substantially to the work~reported.
}

\funding{This work has made use of data from the Asteroid Terrestrial-impact Last Alert System (ATLAS) project. The Asteroid Terrestrial-impact Last Alert System (ATLAS) project is primarily funded to search for near earth asteroids through NASA grants NN12AR55G, 80NSSC18K0284, and 80NSSC18K1575; byproducts of the NEO search include images and catalogs from the survey area. This work was partially funded by Kepler/K2 grant J1944/80NSSC19K0112 and HST GO-15889, and STFC grants ST/T000198/1 and ST/S006109/1. AA acknowledges funding from NSF grant Award Number 1914579, WoU-MMA: Multi-Messenger Studies with Very-High-Energy Gamma Rays.}

\institutionalreview{%In this section, you should add the Institutional Review Board Statement and approval number, if relevant to your study. You might choose to exclude this statement if the study did not require ethical approval. Please note that the Editorial Office might ask you for further information. Please add “The study was conducted in accordance with the Declaration of Helsinki, and approved by the Institutional Review Board (or Ethics Committee) of NAME OF INSTITUTE (protocol code XXX and date of approval).” for studies involving humans. OR “The animal study protocol was approved by the Institutional Review Board (or Ethics Committee) of NAME OF INSTITUTE (protocol code XXX and date of approval).” for studies involving animals. OR “Ethical review and approval were waived for this study due to REASON (please provide a detailed justification).” OR “Not applicable” for studies not involving humans or animals.
}

\informedconsent{%Any research article describing a study involving humans should contain this statement. Please add ``Informed consent was obtained from all subjects involved in the study.'' OR ``Patient consent was waived due to REASON (please provide a detailed justification).'' OR ``Not applicable'' for studies not involving humans. You might also choose to exclude this statement if the study did not involve humans. Written informed consent for publication must be obtained from participating patients who can be identified (including by the patients themselves). Please state ``Written informed consent has been obtained from the patient(s) to publish this paper'' if applicable.
}

\dataavailability{%We encourage all authors of articles published in MDPI journals to share their research data. In this section, please provide details regarding where data supporting reported results can be found, including links to publicly archived datasets analyzed or generated during the study. Where no new data were created, or where data is unavailable due to privacy or ethical restrictions, a statement is still required. Suggested Data Availability Statements are available in section ``MDPI Research Data Policies'' at \url{https://www.mdpi.com/ethics}.
} 

\acknowledgments{The ATLAS science products have been made possible through the contributions of the University of Hawaii Institute for Astronomy, the Queen’s University Belfast, the Space Telescope Science Institute, the South African Astronomical Observatory, and The Millennium Institute of Astrophysics (MAS), Chile.}  %We  combined the acknowledgments section into one paragraph, please check and confirm.}. }

\conflictsofinterest{%Declare conflicts of interest or state ``The authors declare no conflict of interest.'' Authors must identify and declare any personal circumstances or interest that may be perceived as inappropriately influencing the representation or interpretation of reported research results. Any role of the funders in the design of the study; in the collection, analyses or interpretation of data; in the writing of the manuscript; or in the decision to publish the results must be declared in this section. If there is no role, please state ``The funders had no role in the design of the study; in the collection, analyses, or interpretation of data; in the writing of the manuscript; or in the decision to publish the~results''.
} 

%%%%%%%%%%%%%%%%%%%%%%%%%%%%%%%%%%%%%%%%%%
%% Optional

%% Only for journal Encyclopedia
%\entrylink{The Link to this entry published on the encyclopedia platform.}

\begin{adjustwidth}{-\extralength}{0cm}

\reftitle{References}

\PublishersNote{}
\end{adjustwidth}
\end{document}